# Large Fizeau's Light-Dragging Effect in a Moving Electromagnetically Induced Transparent Medium

Pei-Chen Kuan, Chang Huang, Wei Sheng Chan, Sandoko Kosen & Shau-Yu Lan*

Division of Physics and Applied Physics, School of Physical and Mathematical Sciences, Nanyang Technological University, 21 Nanyang Link, Singapore 637371, Singapore

* To whom correspondence should be addressed. Email: sylan@ntu.edu.sg

**As one of the most influential experiments on the development of modern macroscopic theory from Newtonian mechanics to Einstein's special theory of relativity, the phenomenon of light dragging in a moving medium has been discussed and observed extensively in different types of systems. To have a significant dragging effect, the long duration of light travelling in the medium is preferred. Here, we demonstrate a light dragging experiment in an electromagnetically induced transparent cold atomic ensemble and enhance the dragging effect by at least three orders of magnitude compared to the previous experiments. With a large enhancement of the dragging effect, we realize an atom-based velocimeter that has a sensitivity two orders of magnitude higher than the velocity width of the atomic medium used. Such a demonstration could pave the way for motional sensing using the collective state of atoms in a room temperature vapor cell or solid state material.**

The phase velocity of an electromagnetic wave travelling in a moving medium with velocity **v** deviates from the velocity of light in vacuum **c**. Naively, one might expect that in the low speed limit $v<<c$, the phase velocity $\mathbf{v}_p$ can be formulated by the Newtonian velocity addition $\mathbf{v}_p=\mathbf{c}/n+\mathbf{v}$, where $n$ is the index of refraction of the moving medium. The deviation from the Newtonian velocity addition was observed by Fizeau in a flowing water tube experiment[1]. Fresnel predicted the result by making an assumption that ether was partially dragged by the moving medium and adding a dragging coefficient $F_d$ in the velocity addition[2]. The dragging coefficient was later modified by Lorentz including a dispersion term[3]. The phase velocity of the electromagnetic wave can therefore be written as

$$\mathbf{v}_p=\mathbf{c}/n \pm F_d \mathbf{v}, \tag{1}$$

where

$$F_d=1-1/n^2+(\omega/n)[\partial n(\omega)/\partial \omega] \tag{2}$$

and $\omega$ is the angular frequency of the light in the laboratory frame. Few years later, Laue derived the Lorentz dragging coefficient from Einstein's special theory of relativity with relativistic velocity addition to the first order of the medium's velocity[4].

In a non-dispersive medium like water or glass, the dragging coefficient $F_d$ is only on the order of one and therefore a few meters long tube was used in early Fizeau's water tube experiment[1] in order to have an observable effect. Further experiments used a spinning glass rod in a ring resonator to improve the detection sensitivity[5,6]. Although the dragging effect can be enhanced in dispersive atomic vapors when tuning the frequency of light near the atomic resonance, large dispersion is usually accompanied by strong absorption of light. A recent experiment shows the phase velocity dragging in a hot atomic vapor cell by shifting the frequency away from the resonance to avoid the absorption and improves the dragging coefficient $F_d$ by two orders of magnitude[7]. There are proposed experiments using an electromagnetically induced transparent (EIT) medium to enhance the dragging effect for the study of motional sensing, transverse light dragging, and laboratory analog of astronomical systems, such as event horizon in the black hole[8-13]. Group velocity dragging under EIT has been shown in a stationary hot vapor by selecting a group of atoms through optical pumping[14].

In the following, we demonstrate phase velocity dragging in a moving cold $^{85}$Rb atomic ensemble under EIT. An enhancement of the dragging coefficient is achieved by three orders of magnitude compared to the previous experiment[7]. Taking advantage of the large dispersion property of EIT medium, we also show the collective state of atoms can be applied for velocimetry in which the sensitivity is 100 times higher than the Doppler width of the ensemble used.

## Results

**Light dragging medium**. Electromagnetically induced transparency has been studied substantially in both atomic vapors and solid system over the past two decades[15]. Owing to its extraordinary property of slowing down the group velocity of light in a medium without absorption, it finds applications in quantum optics and information science[15-17]. A simple EIT scheme can be implemented in a three-level atomic system, wherein two lower atomic states |g> and |s> with long coherence time are coupled to a third state |e> by optical excitations. A control field resonating on the |s> to |e> transition creates a quantum interference for a probe field resonating on the |g> to |e> transition such that the real and imaginary part of the susceptibility $\chi$ can both approach to zero at the resonance. The slope of the real part of the susceptibility determines the group velocity as can be seen from the index of refraction $n \approx 1+(1/2)\text{Re}[\chi]$ and the group velocity $\mathbf{V}_g=\mathbf{c}/(n+\omega(\partial n(\omega)/\partial\omega))$. The magnitude of group velocity near the resonance can be approximated as $V_g \propto (\Gamma_{ge}\Gamma_{gs}+\Omega_c^2)/N$, where $N$ is the density of atoms, $\Gamma_{ge}$ is the decoherence rate of |g> and |e>, $\Gamma_{gs}$ is the decoherence rate of |g> and |s>, and $\Omega_c$ is Rabi frequency of the control field[15]. The group velocity of the probe field can therefore be reduced by lowering the control field intensity or increasing the atom density. Our three-level system involves $^{85}$Rb D2 line $|g\rangle \equiv |5^2S_{1/2}, F=2\rangle$, $|s\rangle \equiv |5^2S_{1/2}, F=3\rangle$, and $|e\rangle \equiv |5^2P_{3/2}, F'=3\rangle$ as shown in Fig. 1. Figure 2 shows a typical EIT spectrum of our experiment. We fit the spectrum with the transmission $T=\exp(OD\times(\Gamma_{ge}/2)\times\text{Im}[\chi])$ of the probe field[15] and obtain $OD=36$, where $OD=NL3\lambda^2/2\pi$ is the optical depth of the ensemble, $L$ is the length of the ensemble, and $\lambda$ is the wavelength of the probe field. To change the velocity of the center-of-mass motion of the ensemble, we apply a resonant scattering force by imparting a push field on the ensemble. The velocity change is then controlled by the power of the push field.

**Measurement**. Defining the group index of the medium $n_g \equiv c/V_g$ when it is larger than one, the dragging coefficient can be rewritten as $F_d=n_g/n-1/n^2$. The index of refraction $n$ of a medium near the EIT transmission window is approximately unity[15], so the phase velocity can now be further simplified as $\mathbf{v}_p=\mathbf{c}+n_g\mathbf{v}$. To detect the light dragging effect, we compare the phase of the probe field with a local oscillator through the method of heterodyne detection[18]. Considering the phase shift of the probe field passing through a medium with a length $L$ as $\Phi=kL$, where $k$ is the wavevector of the field and propagating along the direction of the moving medium, we can rewrite the phase with the definition of the magnitude of the phase velocity $v_p\equiv\omega/k$ and obtain

$$\Phi=L\omega/v_p=L\omega/(c(1+v/V_g))\approx(L\omega/c)(1-v/V_g) \quad (3)$$

when $v\ll V_g$. By comparing the phase with the local oscillator, we are able to extract the phase shift of light as

$$\varphi=-kLv/V_g=-kLvF_d/c. \quad (4)$$

The phase shift is therefore proportional to the group delay $t=L/V_g$ of the pulse propagating through the medium. The measured light dragging phase with different velocity of atomic cloud is shown in Fig. 3 with control field power of 2 mW and 0.6 mW.

To confirm the light dragging effect, we calculate the expected phase shift using equation 4. The group delay of the probe field is measured at each velocity by fitting the center of the probe field pulse with a Gaussian function. The discrepancy between measured and calculated phase shift is mainly due to the effect of other hyperfine states in the EIT process and also the systematic error of velocity measurement due to the distortion of the atomic cloud after interacting with the pushing field. To take out the extra phase due to the EIT process, we fit our measured phases with a linear function and offset the fitting line to zero when the velocity is at zero. Figure 4 shows the measured delayed phases at the control field power of 0.6 mW and 2 mW and the expected phase delays are in a good agreement within one standard error. With the measured atomic cloud size 1.4 mm and our largest group delay time $t$=855 ns, the dragging coefficient $F_d$ in our experiment has reached $1.83\times10^5$.

## Discussion

Motional sensing using atoms via atomic interference has reached very high precision and accuracy[19,20]. However, due to its nature of differential measurement it can only be sensitive to acceleration. Although two-photon Raman velocimetry can select a group of atoms with very narrow velocity width in an atomic ensemble determined by the duration of the pulse length[21], it is not adequate to sense the collective motion of an atomic cloud. For the determination of the center-of-mass velocity of an atomic ensemble, one would be required to map out all the velocity groups and therefore the sensitivity is restricted to the Doppler broadening of the ensemble. Even in the high precision photon recoil frequency measurement using optical Bloch oscillation with $10^{-9}$ relative uncertainty[22], it can only measures integers of one photon recoil frequency. For light dragging in an EIT medium, all atoms participate to the collective motion so that the velocity measurement is less sensitive to the Doppler broadening of the atomic ensemble. Slow light in a three-level system can also be modeled as a dark state polariton: $\psi=\cos\theta(t')\varepsilon(z,t')-\sin\theta(t')N^{1/2}\sigma(z,t')\exp(ik_{\text{eff}}z)$, where $z$ is the spatial coordinate, $t'$ is time coordinate, $\varepsilon(z,t')$ is the electric field amplitude of probe field, $N^{1/2}\sigma(z,t')$ is the collective atomic spin coherence, and the mixing angle $\theta(t')$ is determined by the coupling strength and control field intensity[23,24]. When the probe field enters the EIT medium, part of the probe field is converted into the collective spin coherence. Due to the motion of the atomic ensemble, the exponent $\exp(ik_{\text{eff}}z)$ can be extended to $\exp[ik_{\text{eff}}(z_o+vt')]$, where $z_o$ is the initial position of the ensemble[25]. After the probe field exits the ensemble, the collective atomic coherence is converted back to the probe field with an additional phase shift $k_{\text{eff}}vt'$, coincides with equation 4 as $t$ is the group delay of the probe field.

Our measured phase uncertainty is about 0.01 radians by taking the mean of three cycles of 70 MHz sinusoidal wave in the probe field envelope and averaging for 20 experimental cycles. Each experimental cycle takes 2 s and the duration is mainly limited by the time of loading the atomic ensemble and processing the data. Using the value of the effective wavevector $k_{eff}$=1.61×10$^7$ m$^{-1}$ and largest group delay time $t$=855(7) ns, our experiment demonstrates a velocimeter with sensitivity $\Delta v=\Delta\varphi/(k_{eff}t)$ at the level of 1 mm s$^{-1}$, two orders of magnitude higher than the velocity width $\Delta v_a=(8k_BT\ln2/m)^{1/2}\approx$176 mm s$^{-1}$ of our atomic ensemble, where $k_B$ is the Boltzmann constant, $m$ is the mass of $^{85}$Rb, and $T$ is the effective temperature of the atomic cloud. In principle, with our 1 µW of probe field, we should be able to increase the sensitivity by at least two orders of magnitude when we reach the shot noise limit at 5×10$^{-4}$ radians per square root Hertz by recording all the cycles within the probe field. The sensitivity can also be improved by using larger atomic ensemble and smaller group velocity, i.e. 1 cm of an atomic sample can improve our sensitivity to 100 µm s$^{-1}$. Storage of the optical field in an atomic ensemble has reached a storage time close to a minute by either confining cold atomic vapor in an optical potential or placing a rare earth-ion-doped crystal at cryogenic temperature[26,27]. The sensitivity can be improved by seven orders of magnitude with successful implementation of the above methods. To measure the gravity with our velocimeter, equation 4 can be expressed as $\varphi$=-$k_{eff}gt^2$. One second of the storage time can induce a phase shift of 10$^8$ radians, reaching the level of the current state-of-the-art phase shift of atom interferometer based inertial sensor[28,29].

In conclusion, we have demonstrated the largest Fizeau's light dragging effect using a moving EIT medium and applied it for velocimetry. Tracing the velocity of a free-falling atomic ensemble at different timing, one can measure the acceleration as well. Although the counter-propagating arrangement of the EIT fields in our experiment can only be implemented with cold atoms due to Doppler broadening of the ensemble[15], this method can be extended to thermal atoms by using co-propagating arrangement which is insensitive to the Doppler broadening of atoms to the first order. Our demonstration could lead to the study of inertial effect with a collective state of atoms and designing a new type of motional sensor.

## Methods

**Derivation of the dragging coefficient.** Consider a probe field travelling along a moving medium of velocity $v$, the dispersion relation in the rest frame reads

$$k' = n(\omega')\omega'/c, \tag{5}$$

where $k'$ and $\omega'$ are the wavenumber and the frequency of the probe field in the rest frame. Employing the Lorentz transformation to the first order of $v/c$, $\omega' = \omega - kv$, $k' = k - \omega v/c^2$, where $k$ and $\omega$ are the wavenumber and frequency of the field in the laboratory frame, we expand the index of refraction $n(\omega')$ in equation 5 in a power series of $kv$ to the first order

$$k \approx n(\omega)\omega/c - vk\partial(n(\omega)\omega/c)/\partial\omega + \omega v/c^2. \tag{6}$$

Dividing equation 6 by $n/(ck)$, the phase velocity $v_p \equiv \omega/k$ can be written as $v_p = c/n + F_d v$, where $F_d = 1 - 1/n^2 + (\omega/n)(\partial n(\omega)/\partial\omega)$ is the dragging coefficient.

**Experimental details.** Our medium is an ensemble of about $10^9$ $^{85}$Rb atoms after loading from a magneto-optical trap (MOT) and the effective temperature is about 40 μK after sub-Doppler cooling. Due to the imbalance of radiation pressure from the cooling beams and gravity, the atomic cloud has an initial velocity before the EIT fields are sent in. Our push field is resonating on the $^{85}$Rb D2 line $F=2$ to $F'=3$ transition to the ensemble aided by an optical pumping field resonating on $F=3$ to $F'=2$ of D1 line to ensure atoms are returned to the original state. The pulse duration of the push field is 0.7 ms and the power is adjusted for varying the velocity. Atoms absorb photons from the push field in a well-defined direction and re-scatter them in a random direction. On average, atoms will gain a velocity proportional to the number of the photons being absorbed. The direction of the push field can be reversed for measurements of velocity at the opposite direction. The velocity of the atomic cloud after the push field is measured using the time-of-flight method with a CCD camera.

The probe field has a waist of 300 μm positioned around the center of the atomic ensemble and the waist of the control field is about two times larger than the probe beam to ensure all the atoms interacting with the probe field are addressed by the control field with the same intensity. We align the control and the probe field at nearly counter-propagating direction (about 183 degrees). The wavevector $k$ in equation 4 can be replaced by the effective wavevector $k_{\text{eff}}=k-k\cos 183^{\circ}$. The control field is generated from a diode laser and part of the power is sent through an electro-optical modulator. The first sideband after the modulator passes through a solid Fabry-Pérot cavity followed by a 70 MHz acoustic-optical modulator (AOM). The field coming out of lower first order serves as the probe field and the zero order serves as an auxiliary field which then combines with the probe field by a polarizing beam splitter to form a 70 MHz beating signal. This 70 MHz signal is further split: part of the beam is sent through the atomic ensemble for the light dragging experiment and the other half serves as a local oscillator for phase comparison as shown in Fig. 1b. Since the auxiliary field is 70 MHz detuned from the probe field, it does not experience the large light dragging effect as the probe field and therefore the phase shift of the 70 MHz signal results from the phase velocity dragging of the probe field only.

After 5 ms of turning off the magneto-optical trap, the push field is on followed by probe and control field. The probe field intensity is about 1 μW and its amplitude is modulated by a Gaussian function of 9 μs full width at half maximum. The control field is turned on 300 μs before the probe field to ensure atoms are prepared in the $F=2$ state.

## Acknowledgments

We thank David Wilkowski, Holger Müller, and Dzmitry Matsukevich for stimulating discussions. This work is supported by Singapore National Research Foundation under Grant No. NRFF2013-12, Nanyang Technological University under start-up grants, and Singapore Ministry of Education under Grants No. Tier 1 RG193/14.

## Author Contributions

S.L. and P.K. conceived and designed the experiment. S.L. wrote the manuscript. S.K. and W.C contributed to the early stage of the experiment. P.K. and H.C. took and analyzed the data. All authors contributed to the experimental setup, discussed the results and commented on the manuscript.

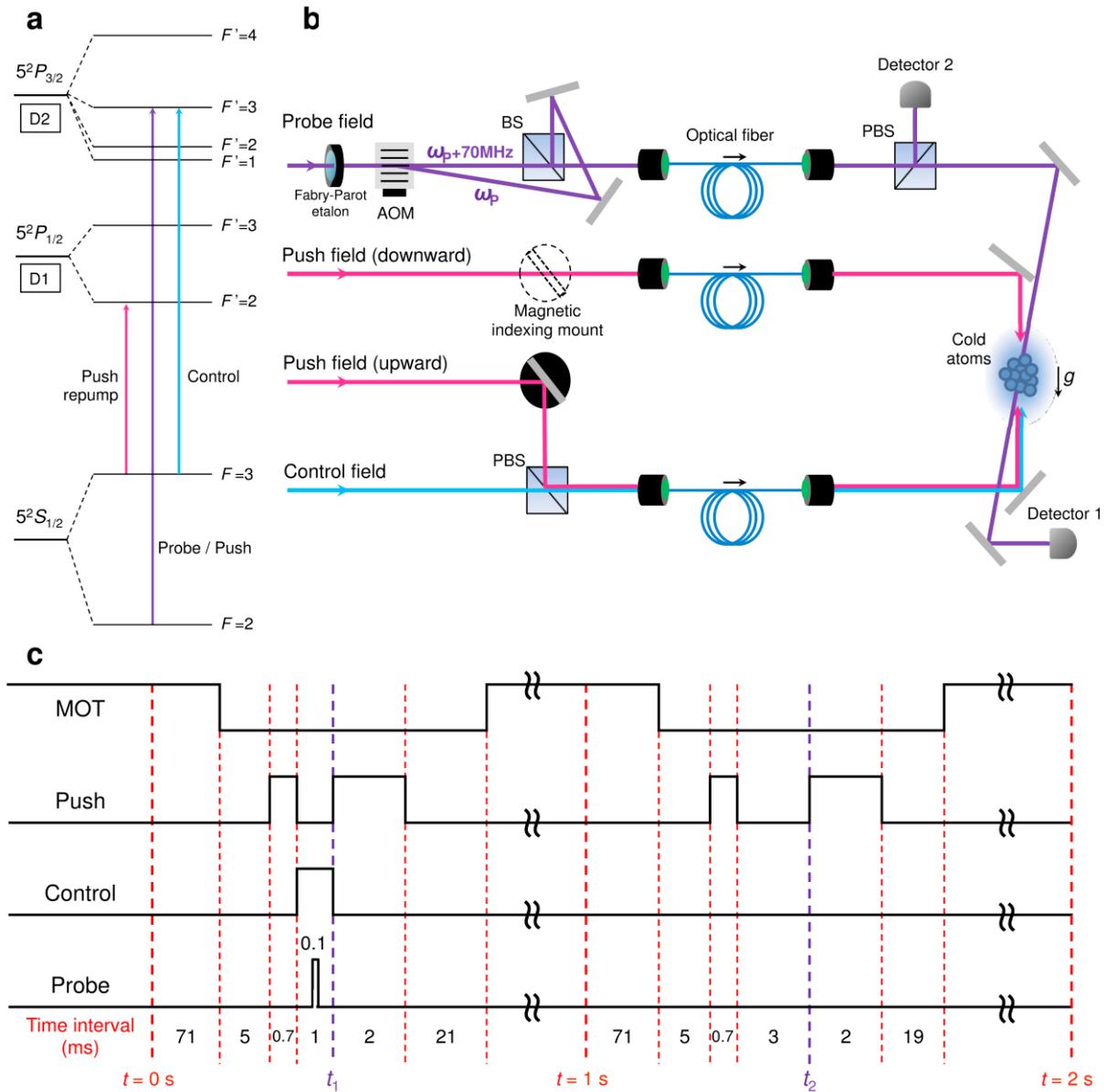

**Figure 1 | Experimental Details.** (**a**) Level diagram of relevant atomic transitions for the experiment. (**b**) Experimental setup: BS is a beam splitter. PBS is a polarizing beam splitter. AOM is an acoustic-optical modulator. We use a magnetic indexing mount to switch between pushing atoms upward and downward. In the pushing upward setup, the push field is overlapped with the control field and coupled to a single mode fiber. In the push downward case, the push field is coupled backward to the control field fiber exiting port with 75% coupling efficiency to ensure the overlap of the control and push field. Detector 2 records the reference field for comparing the phase of the probe field from detector 1 on an oscilloscope. The local gravity $g$ is pointing downward. The probe field frequency is $\omega_\text{p}$. The probe and control fields are aligned around 183°. (**c**) Timing sequence of the experiment. Magneto-optical trap (MOT) represents the timing sequence for cooling and repumping fields as well as the magnetic field for the preparation of the cold atomic ensemble. The push field is on at $t=t_1$ and $t=t_2$ for the determination of the velocity by imaging the position of the atomic cloud 1 ms and 3 ms after the push field.

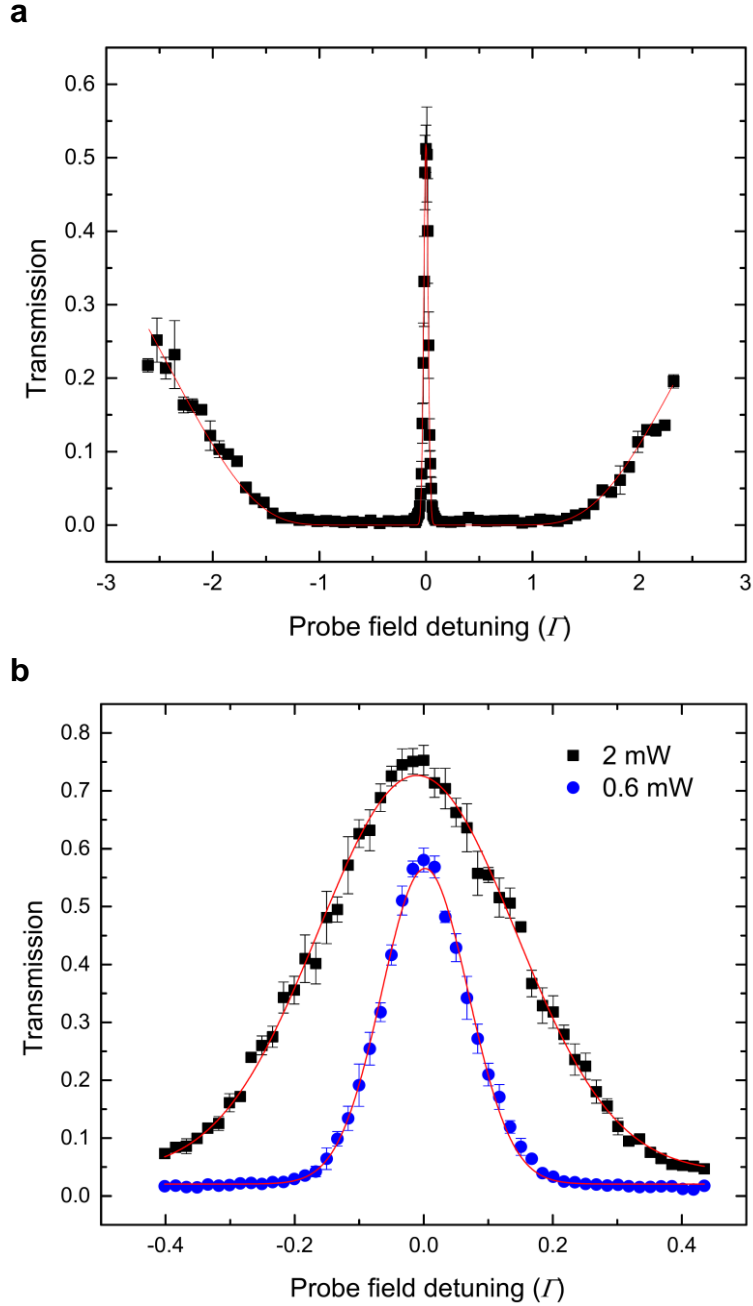

**Figure 2 | Transmission of the probe field versus detuning.** The probe field is on $5^2S_{1/2}$, $F=2$ to $5^2P_{3/2}$, $F'=3$ transition and the control field is on $5^2S_{1/2}$, $F=3$ to $5^2P_{3/2}$, $F'=3$ transition. The probe field detuning is expressed in terms of excited state spontaneous decay rate $\Gamma$. The standard error of each data point is calculated from the average of three experimental trials. (**a**) Electromagnetically induced transparency spectrum for optical depth (*OD*) measurement. The fitting curve shows *OD*=36.0(0.4) and control field Rabi frequency $\Omega_c$=0.582(5)$\Gamma$. The duration of the probe field is 100 μS and the control field is turned on 100 μS before the probe field is on in order to prepare the state of atoms in the $5^2S_{1/2}$, $F=2$. The standard error of the central peak is calculated from five data points while the rest are two data points. (**b**) Transmission peak of probe field with 2 mW (black squares) and 0.6 mW (blue circles) control field power. The fitted Gaussian functions of black squares an blue circles give $1/e^2$ width of 0.306(6)$\Gamma$ and 0.134(1)$\Gamma$, respectively. The standard error of each data point is calculated from the average of three experimental trials.

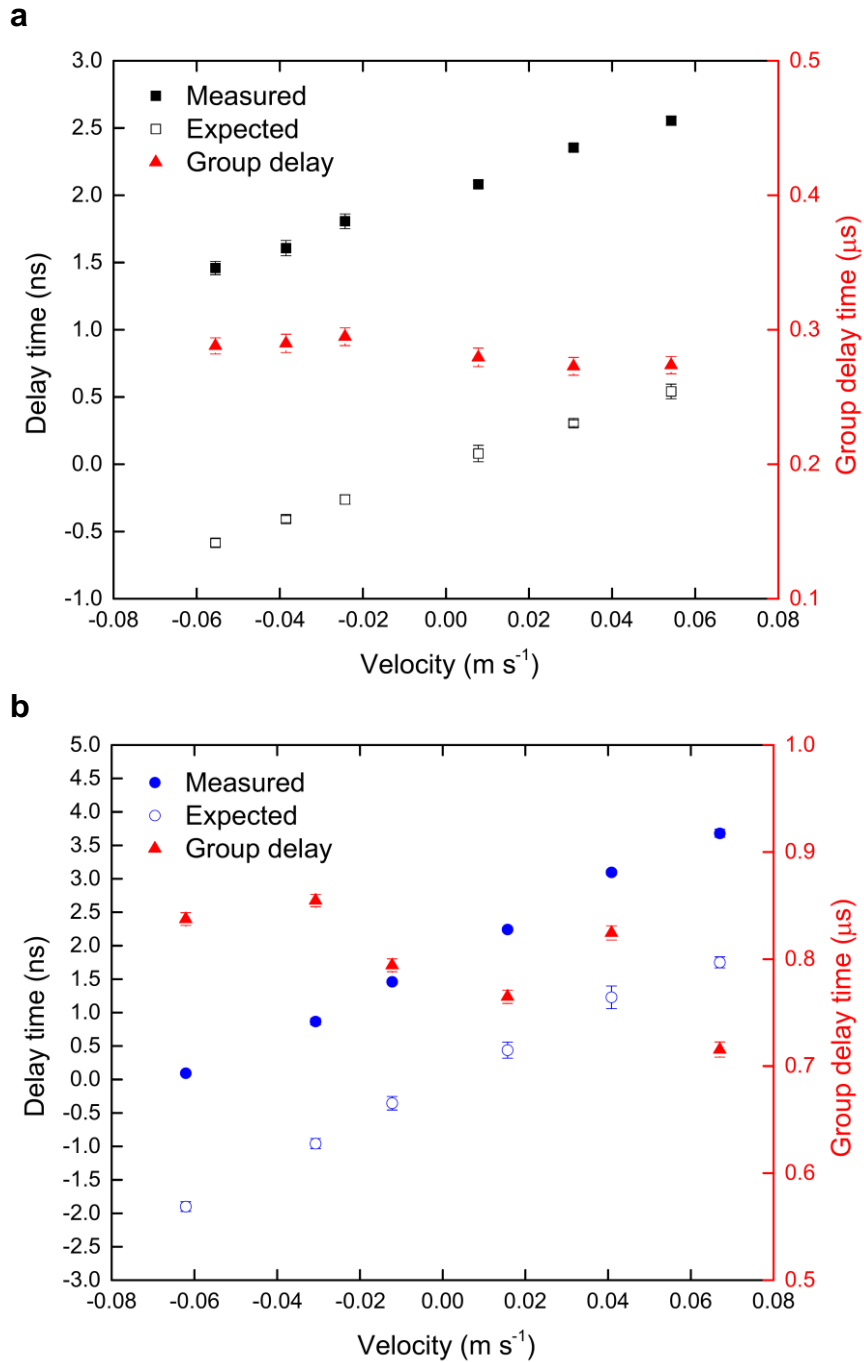

**Figure 3 | Phase and group delays of the probe field versus velocity of the atomic ensemble.** (**a**) With control field power of 2 mW. The black solid squares are the measured phases and the black open squares are the expected phases (left axis) from equation 4 and group delay measurements. (**b**) With control power of 0.6 mW. The blue solid circles are the measured phases and the blue open circles are the expected phases (left axis) from equation 4 and group delay measurements. The red solid triangles are the group delay times (right axis). The phase delay are measured in terms of the delay time. One cycle corresponds to 1/70 MHz=14.29 ns. The measured phase uncertainty is by taking the standard error of three cycles of 70 MHz sinusoidal wave in the probe field envelope and averaging for 20 experimental cycles. Each experimental cycle takes 2 seconds. The group delay of the probe field is measured at each velocity by fitting the center of the probe field pulse with a Gaussian function.

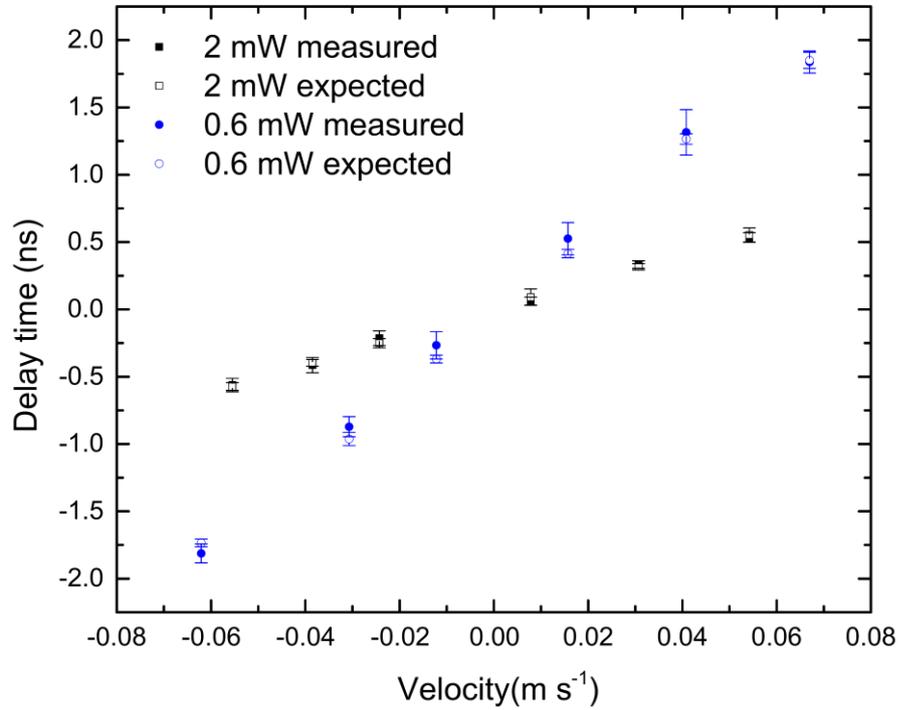

**Figure 4 | Phase delay of the probe field with offset versus velocity of the atomic ensemble.**
The phase delay times from Fig.3 are offset to zero at zero velocity. The solid circles (0.6 mW control field power) and squares (2 mW control field power) are the measured delayed phases and the open circles (0.6 mW control field power) and squares (2 mW control field power) are the expected delayed phases from equation 4 and group delay measurements.